\documentclass[10pt,letterpaper]{article}
\usepackage{opex3}
\usepackage{graphicx}
\usepackage{dcolumn}
\usepackage{bm}
\usepackage[normalem]{ulem}	
\usepackage{xcolor}

\usepackage{ifthen}
\usepackage[mathscr]{euscript}
\usepackage{mathtools}
\usepackage{amsmath,amssymb,amsthm}
\usepackage{bbold}

\begin{document}

\title{Absolute calibration of fiber-coupled single-photon detector}

\author{Tommaso Lunghi, Boris Korzh, Bruno Sanguinetti and Hugo Zbinden}

\address{Group of Applied Physics, University of Geneva, Chemin de Pinchat 22, CH-1211 Gen\`eve 4, Switzerland}

\email{Tommaso.Lunghi@unige.ch} 

\begin{abstract}
We show a setup for characterising the efficiency of a single-photon-detector absolutely and with a precision better of 1\%. Since the setup does not rely on calibrated devices and can be implemented with standard-optic components, it can be realised in any laboratory. Our approach is based on an Erbium-Doped-Fiber-Amplifier (EDFA) radiometer as a primary measurement standard for optical power, and on an ultra-stable source of spontaneous emission. As a proof of principle, we characterise the efficiency of an InGaAs/InP single-photon detector. We verified the correctness of the characterisation with independent measurements. In particular, the measurement of the optical power made with the EDFA radiometer has been compared to that of the Swiss Federal Office of Metrology using a transfer power meter.
Our approach is suitable for frequent characterisations of high-efficient single-photon detectors.
\end{abstract}


\section{Introduction}
Recently, important developments have been achieved in single-photon-counting technologies and high-efficient detectors have been developed in several laboratories\,\cite{RestelliAPL13,marsili2013detecting, Miki:13}. 
These improvements allow us to perform challenging experiments in quantum optics, such as Device-Independent quantum-key distribution\,\cite{art:Gisin10} and detection-loophole-free Bell tests\,\cite{art:Pearle70,art:Christensen13}. However, while the system detection efficiency is rapidly increasing, an accurate characterisation method is still not easy to access.

Conventionally, calibration of a single-photon detector is obtained by measuring the power of a classical beam with a reference power meter, then the beam is strongly attenuated and sent to the detector under test, DUT\,\cite{art:Cheung:11}. The accuracy of this method (usually between 5\% and 10\%) is mainly limited by the power stability of the beam, the precision of the attenuation stage and accuracy of the reference power meter. 
In particular, the accuracy of the absolute measurement for commercial power meters is large ($\sim$\,5\,\%) so a direct calibration traceable to the primary standard is required. This calibration can only be performed in a metrological laboratory having the primary measurement standard, e.g. the cryogenic substitution radiometer\,\cite{art:Parr00}. This process is long and time-consuming. 

An alternative approach has been developed on the basis of correlated photon pairs\,\cite{art:Polykov07, art:Polykov09}. It allows us to estimate the detection efficiency of the DUT without relying on calibrated devices. In this scheme, the fraction of emitted photons impinging on the DUT has to be determined precisely. This is challenging due to the many spatial modes involved, particularly for fiber-coupled detectors\,\cite{art:Ware04}. Therefore, the conventional method is still commonly employed.

Sanguinetti {\it et al.}\,\cite{art:Sanuinetti2012} demonstrated an absolute measurement of classical light at 1540\,nm using a radiometer based on an Erbium-Doped Fiber Amplifier (EDFA). This radiometer is appealing for two reasons: it can be built with standard optical components and it is suitable for low powers in the range of nW, allowing a weaker attenuation stage. 

Moreover, when characterising single-photon detectors based on semiconductor materials (SPAD), non-linearities such as afterpulsing and dead time/hold-off time have to be corrected in order to reduce errors. Time-correlated detection-efficiency measurements\,\cite{art:Lunghi12} attenuate the impact of these non-linearities at the expense of a more complicated setup. A simpler but efficient approach is thus welcome.

In this paper we present a testbench for absolute characterisations of single-photon detectors with high precision. Our setup is based on the conventional method, but we avoid the calibration at the metrological laboratory by using an EDFA radiometer that has been built following\,\cite{art:Sanuinetti2012}. We calibrate the reference power meter with this device and then we verify the calibration factor at the Swiss Federal Office of Metrology. Compared to\,\cite{art:Sanuinetti2012}, the calibration has been performed at a different wavelength (1552\,nm) and includes a full error analysis.   
To improve stability and repeatability, we have developed a stable light source of spontaneous emission\,\cite{art:Monteiro2013}. This source has been demonstrated as a precise reference for metrological applications since it is stable over days, unpolarised and incoherent.
As a proof of principle, we characterise the efficiency of a SPAD analysing extensively the uncertainty on the efficiency measurement and compensating the impact of non-linearities.\\
This paper aims to provide a rigorous and reliable calibration protocol for characterising the efficiency of a SPAD. It is organized as follows: first of all we give an overview of the actual implementation devoting particular attention to the EDFA radiometer and the optical source. Then we show how to characterise the uncertainty introduced by each component. Particular attention is given to the calibration of the reference power meter traceable to the EDFA radiometer. Finally, we show how to characterise the detection efficiency of a SPAD.

\section{Experimental setup}
\subsection{Overview}
\begin{figure}[!htbp]
\centering\includegraphics[width=6cm]{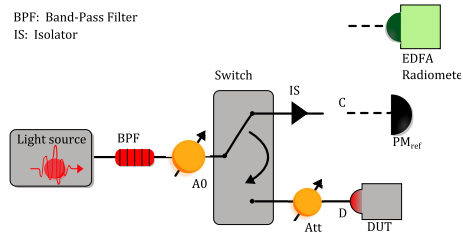}
\caption{Conventional method: sketch.} 
\label{fig:overview_setup}
\end{figure}

Figure\,\ref{fig:overview_setup} shows the testbench we use for the efficiency characterisation. Here we give an overview of the proposed testbench while the implementation is detailed in Sec.\ref{sec:setup}. The beam emitted by the source is sent with an optical switch, either to the reference power meter (PM$_{\text{ref}}$) or to the DUT. The power in the beam is attenuated in two stages, $A_0$ and $Att$, before impinging on the DUT. 
During the measurement PM$_{\text{ref}}$ monitors periodically the beam power. When calibrating PM$_{\text{ref}}$, the light is also sent to the EDFA radiometer. 

\subsection{Incoherent stable source of spontaneous emission}\label{sec:Source}

An appropriate light source is vital for the stability and repeatability of the measurement. Although laser sources have the advantage of well-defined wavelength, the long coherence of the emitted light degrades the repeatability of the calibration because of interreflections that can occur between any couple of semi-reflecting surfaces\,\cite{art:Envall04}. Moreover, polarisation-dependent losses degrade the stability of the entire setup. Commercial incoherent sources are good alternatives, however their power may deviate by more than 2000 ppm after a few hours of measurement\,\cite{EXFO}. 

\begin{figure}[!htbp]
\centering\includegraphics[width=7cm]{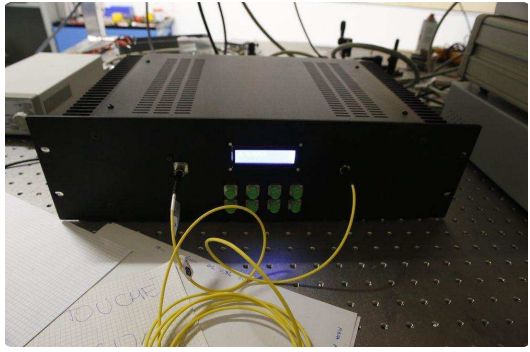}
\caption{
Picture of the stand-alone device: the output fiber (on the right) is plugged into the power meter (on the left).} 
\label{fig:sourcepicture}
\end{figure}

Another possible alternative is represented by an inverted atomic medium, such as Erbium-Doped Fibre (EDF)\,\cite{art:Monteiro2013}.
In this system, a strong 982\,nm pump laser promotes all the Erbium ions in a metastable, excited state. When an ion decays to its ground level a photon is emitted. When the emission coming from the EDF is dominated by the spontaneous component, the light guided in the fibre is unpolarised and has a short coherence time as it can be deduced by the broad spectrum (centred at 1530\,nm and 48\,nm broad). Moreover, under strong pumping, the output optical power saturates with extremely small variations.
Monteiro {\it et al.}\,\cite{art:Monteiro2013} reported less than 25 \,ppm of power deviation after 3 days of measurement for a short-length EDF.\\
Based on their work, we have developed a stand-alone device (see Fig.\ref{fig:sourcepicture}) increasing the length of the EDF (ER30-4/125 by Thorlabs) to $\sim$18\,cm (emitted power $\sim\mu$W).
The relative Allan deviation of the output power shows $\leq$20 ppm of deviation after one day of measurement. 

\subsection{EDFA Radiometer: operating principle}
As mentioned previously, the accuracy of the conventional method relies on the accuracy of the reference power meter. 
When measuring a light beam with an uncalibrated but linear power meter, the reading will be off by a factor $k$ compared to the real value. Absolute optical power measurements can be obtained with the EDFA radiometer.\\ 
Like the stable source, this device is based on the EDF: the number of output photons per mode $\mu_{out}$ exiting from an EDF depends on stimulated and spontaneous emission and is described by\,\cite{art:Sanuinetti2012}:
\begin{equation}\label{eq:photonpermode}
\mu_{out} = G\mu_{in} + G - 1
\end{equation}
where $\mu_{in}$ is the number of input photons per mode, and $G$ is the gain of the medium.
The term $G\mu_{in}$ represents the emission stimulated by the input light while the term $G-1$ represents the spontaneous emission.
Using the formalism of Eq.\,\ref{eq:photonpermode}, the measured optical powers exiting the fibre when we inject ($P_{st}^*$) or not ($P_{sp}^*$) an input light are given by
\begin{eqnarray}\label{eq:power1}
P_{sp}^*=(G - 1)\cdot\frac{2}{\tau_c}\cdot h\nu \cdot k\\
P_{st}^*=(G\mu_{in} + G - 1)\cdot\frac{2}{\tau_c}\cdot h\nu\cdot k\label{eq:power2}
\end{eqnarray}
where h$\nu$ is the photon energy and $2/\tau_c$ is the number of modes per second ($1/\tau_c$ is the number of temporal modes and the factor 2 corresponds to the number of polarisation modes). 
$\mu_{in}$ can be derived from Eq.\ref{eq:power1},\ref{eq:power2}\,\cite{art:Sanguinetti10}: 
\begin{equation}\label{eq:muin}
\mu_{in}(\lambda) = \left(1-\frac{1}{G(\lambda)}\right)\left(\frac{P_{st}^*(\lambda)}{P_{sp}^*(\lambda)}-1\right)
\end{equation}
We stress that $\mu_{in}$ is an absolute measurement since $P_{st}^*(\lambda)/P_{sp}^*(\lambda)$ does not depend on $k$. The gain $G(\lambda)$ can always be deduced independently of $k$ using:
\begin{equation}\label{eq:G}
G(\lambda) = \frac{P_{st}^*(\lambda)-P_{sp}^*(\lambda)}{P^*_{in}(\lambda)}
\end{equation}
where $P^*_{in}(\lambda)$ is the reading of the incoming power. The $\lambda$ dependencies introduced in Eq.\ref{eq:muin}\,,\ref{eq:G} are meant to emphasise that $\mu_{in}$ is obtained for each temporal/spectral mode. 

\begin{figure}[!htbp]
\centering\includegraphics[width=8cm]{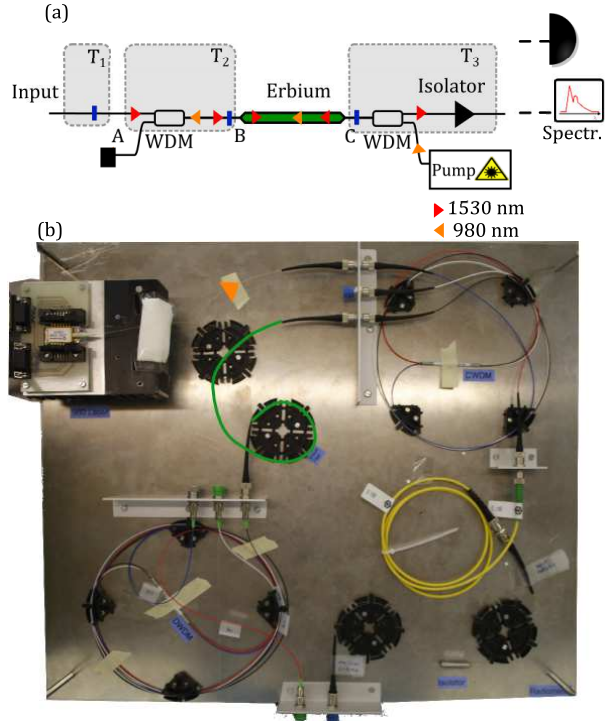}
\caption{(a) Schematic of the radiometer. (b) Picture of the setup. The green line (orange arrow) enhances the Er$^{3+}$ fiber (the pump laser fibre).} 
\label{fig:setupradiometer}
\end{figure}

Following\,\,\cite{art:Sanuinetti2012}, we have built the fibred setup shown in Fig.\ref{fig:setupradiometer}(a). 
The pump light is first injected into the EDF using a WDM, then it is removed with another WDM. We used a counter-propagating pumping configuration to prevent the pump light from exiting at the output of the radiometer. The input light travels through the two DWMs and the EDF where it is amplified. At the output the light is sent either to a spectrometer (Anritsu, MS9710A) or to a power meter. Two isolators are introduced at the input and output of the device to suppress any back-reflection of the telecom light that can occur outside the radiometer. Within the radiometer, back-reflections are suppressed using APC-connectors. 

\section{Absolute calibration of the reference power meter}\label{SEC:EDFA}

\subsection{EDFA radiometer characterisation}

To correctly determine the number of incoming photons, the transmission factors before and after the EDF have to be carefully measured. %
We label the transmission of the input connector as {\it T$_1$}, {\it T$_2$} is the transmission up to the EDF, and {\it T$_3$} is the output transmission (see Fig.\ref{fig:setupradiometer}). The pump laser is off. When connecting an optical fibre to the power meter, the reading can be affected by a systematic calibration error that can be as high as 10\%\,\,\cite{art:Envall04}. To reduce the impact of this error we employ an integrating sphere photodiode (Thorlabs, S144C). This photodiode, however, has a poor power sensitivity ($\geq$1\,$\mu$W) therefore we perform this measurement using a commercial amplified spontaneous emission source (Trillium Photonics) as the input source. The impact of the statistical errors is reduced by repeating several times the power measurements. 
To correctly calibrate {\it T$_1$} we connect and disconnect repeatedly the input fibre into the input connector maximising the transmission each time in order to improve the repeatability\,\cite{art:Sanuinetti2012}. 
The uncertainty on {\it T$_1$} can be estimated from the statistics of the measured transmission factors. {\it T$_2$}, instead, is calibrated by injecting the light backwards into the radiometer and measuring the light firstly in B, then in A. All the connectors belonging to {\it T$_2$} are adjusted to maximise the transmission. In this way, T$_2$ is affected only by statistical errors which scale with the square root of the number of acquisitions. Finally, we calibrate $T_3$ by injecting light from the input port and measuring the light in C and then at the output. Again, all the connectors belonging to $T_3$ are adjusted in order to maximise the transmission. Also in this case the uncertainty scales with the square root of the number of acquisitions. 
The results are tabulated in Tab.\ref{tab:transmissionfactors}:
\begin{table}[!h]
\begin{center}
\begin{tabular}{|c|c|c|}
\hline
\multicolumn{3}{|c|}{\bf{Transmission factors}}\\
\hline
\bf{Sym.}&\bf{Value}& \bf{Rel. uncertainty (k=1)}\\
\hline
T$_1$& 0.967(2)& 0.2\%\\
\hline
T$_2$& 0.8788(9)& 0.1\%\\
\hline
T$_3$& 0.5430(5) & 0.09\%\\
\hline
\end{tabular}
\end{center}
\caption{Transmission factors. {\bf k} indicates the coverage factor.} 
\label{tab:transmissionfactors}
\end{table}
\newline Once the transmission losses are characterised the radiometer is connected to the output C of the optical switch (see Fig.\ref{fig:overview_setup}).

\subsection{EDFA radiometer: gain measurement}
The last quantity to be characterised is the gain of the EDF. 
Using the reference power meter (Thorlabs, S154C) we monitor the light at the output of the radiometer by injecting or not an input beam. In the first case, the reading of the power meter corresponds to both spontaneous and stimulated emission ($P^*_{st}$), while in the second situation it is only due to the spontaneous component ($P^*_{sp}$).
The input light is blocked using A$_{0}$ (EXFO, FVA-3150 equipped with an optical shutter). After having measured the light at the input of the radiometer always with the reference power meter ($P^*_{in}$), we can calculate $G$ using
\begin{eqnarray}\label{eq:gaincorrected}
\overline{G}=\frac{P^*_{st}-P^*_{sp}}{T_3P^*_{in}T_1T_2}\\
\sigma_G = \sqrt{\sigma_{P^*_{st}}^2+\sigma_{P^*_{sp}}^2+\sigma_{P^*_{in}}^2+\sigma_{T_1}^2+\sigma_{T_2}^2+\sigma_{T_3}^2}
\end{eqnarray}
and obtaining $\overline{G} = 6.67(2)$. Note, in Eq.\ref{eq:gaincorrected}, all powers must be read using the same $k$, so we take care to use the same power-measurement-range setting during the three measurements.
For simplicity, we introduce $\overline{G}$ corresponding to the average gain over the spectrum of the input light. In the actual measurement we calculate the gain for each wavelength as described in\,\cite{art:Sanuinetti2012}and in Appendix\,.\ref{SM:radio}. %
 
\subsection{Absolute measurement of the input power}
The input power is deduced absolutely, by measuring first the spontaneous emission from the EDF ($P_{sp}^*$), then all the output light when the input beam is injected into the EDF ($P_{st}^*$). Those powers are recorded with a spectrometer (Anritsu, MS9710A) to recover the dependencies on $\lambda$. With Eq.\ref{eq:muin} we obtain $\mu_{in}(\lambda)$. We convert this number into optical power ($P_{\text{radio}}$) using\,\cite{art:Sanuinetti2012} 
\begin{equation}\label{eq:Pradio}
P_{\text{radio}}=\frac{1}{T_1T_2}\int \mu_{in}(\lambda)\cdot \frac{2 h c^2\Delta\lambda}{\lambda^3}d\lambda
\end{equation}
where $\lambda$, $\Delta\lambda$ are given by the spectrometer employed in the measurement. 
By this measurement we can calibrate PM$_{\text{ref}}$. We define $k_{\text{PM}}$ as the ratio between the power measured with the radiometer and the reading of the reference power meter, i.e. $k_{\text{PM}}=P_{radio}/P_{in}^*$. When calculating the effective amount of light measured with the power meter, one must take into account that a disconnected APC fibre end has an output loss of 3.3\%, due to Fresnel reflection\,\cite{fused_silica}.

We estimate the uncertainty on $k_\text{PM}$ for a single-shot measurement. We detail the analysis in the Appendix\,\ref{SM:radio} obtaining 0.6\% of relative standard error. To test the repeatability of the measurement, we repeat the entire characterisation described in Sec.\ref{SEC:EDFA} three times. The average between the obtained $k_\text{PM}$s yields 95.2\% with a standard deviation of 0.4\%. This is consistent with the uncertainty of the single-shot measurement. As a confirmation, the calibration factor of PM$_{\text{ref}}$ has been measured at the Swiss Federal Office of Metrology (METAS) at 1550\,nm against a transfer power meter traceable to the cryogenic radiometer. In this case, the absolute-calibration factor is measured at 100$\mu$\,W yielding k$_{ABS,Metas}$=95.3$\pm$0.7\%. Considering the non-linearity coefficient between 100$\mu$\,W\, and \,10\,nW (measured at 1541\,nm with a Ge photodiode, k$_{NL,Metas,10nW}$=99.754$\pm$0.001\%), k$_{METAS}$ =95.1$\pm$0.7\%.
We stress that the measurement with the radiometer was a \textit{blind} test since METAS characterised the reference power meter only after we have measured $k_\text{PM}$.
The two values agree within less than one standard deviation, confirming the potential of the radiometer as a measurement device for the primary standard.

\section{Characterisation of the measurement setup}
\subsection{Testbench: actual implementation}\label{sec:setup}

Once the reference power meter is calibrated, we proceed in characterising the efficiency measurement setup. Here we describe in details the testbench shown in Fig.\ref{fig:overview_setup}: the light emitted by the source is filtered at $\lambda_t$\,=\,1552$\pm$1.67\,nm (DiCon, TF500). The beam passes through A$_0$ (EXFO, FVA-3150) and goes towards the optical switch (Lightech, LT-210) that diverts the light either to PM$_{\text{ref}}$ (Thorlabs, S154C) or to the DUT. $Att$ consists of two variable attenuators, A$_1$,A$_2$ (EXFO, FVA-60b). All the APC connectors have been polished with the same equipment to guarantee the same angle. The number of incoming photons per second on the detector, $N$, is derived from the measurement of the light power made with PM$_{\text{ref}}$ using
\begin{equation}
N = \frac{P_{\text{PM}}\cdot R_{DC}\cdot Att}{hc/\lambda_t}
\end{equation}
where $P_{\text{PM}}$ is the power measured by PM$_{\text{ref}}$, $hc/\lambda_t =1.28\cdot 10^{-19}$\,J is the photon energy, $Att=A_{1}\cdot A_{2}$ and $R_{DC}$ is the splitting ratio.

\subsection{Splitting ratio}
The repeatability and stability of the optical switch and the stability of $Att$ affect the uncertainty of $N$. After setting $Att$ to the minimum value, we characterise the splitting ratio, defined as R$_{DC}$=$\frac{P_D}{P_C}$ where P$_C$ (P$_D$) is the power measured at the point C (point D). We measure the light firstly in the upper path then in the bottom path over an extended period ($\sim$10 hours). We use PM$_{\text{ref}}$ to monitor the light at C (see Fig.\ref{fig:overview_setup}) and an EXFO PM1100 power meter calibrated against PM$_{\text{ref}}$ to monitor the light at D. Each power measurement is averaged for 1 second and repeated 10 times on each path. We obtain R$_{DC}$=0.3441\,(3) which corresponds to a relative standard error of 0.09\%.

\subsection{Attenuation stage}
On the basis of the detector characteristics, we choose N$\sim$20000 ph/s to guarantee enough detections without attaining saturation. Given $P_{\text{PM}}$=9\,nW we set A$_{1}$=A$_{2}$= 30\,dB. The remaining attenuation is introduced adjusting A$_{0}$.
To characterise the variable attenuator over 30 dB of attenuation we can not use the stable source since its output power is low, so we use the commercial source of amplified spontaneous emission, filtered at $\lambda_t$\,=\,1552$\pm$1.67\,nm again. A$_i$ is obtained measuring the incoming light when the attenuation is set to the minimum ($\mu$\,W range) and when the attenuation is increased by 30\,dB. Each power measurement is averaged over 1\,s and repeated 10 times. The attenuation value is 
\begin{equation}
A_i= \frac{P_{\text{30dBm}}k_{\text{NL30dBm}}}{P_{\text{60dBm}}k_{\text{NL60dBm}}}
\end{equation}
where $k_{NLrange}$ is the linearity measurement of the reference power meter performed at METAS. 
The standard error is 
\begin{equation}
\sigma_{A_i} = \sqrt{\sigma_{P_{\text{30dBm}}}^2+\sigma_{P_{\text{60dBm}}}^2+\sigma_{\text{NL30dBm}}^2+\sigma_{\text{NL60dBm}}^2} 
\end{equation}
From that we can deduce the standard error on $Att$
\begin{equation}
\sigma_{Att} = \sqrt{\sigma_{A_1}^2+\sigma_{A_2}^2}\sim\sqrt{2}\sigma_{A_i}
\end{equation}

\begin{table}[!h]
\begin{center}
\begin{tabular}{|c|c|c|}
\hline
\multicolumn{3}{|c|}{\bf{Transmission Factors}}\\
\hline
\bf{Sym.}&\bf{Value}& \bf{Rel. uncertainty (k=1)}\\
\hline
A$_1$& 0.99690(96)$\cdot$10$^{-3}$& 0.096\%\\
\hline
A$_2$& 0.97203(93)$\cdot$10$^{-3}$& 0.096\%\\
\hline
k$_{NL30dBm}$& 0.998203(9)& 0.0009\%\\
\hline
k$_{NL60dBm}$& 0.99671(2) & 0.002\%\\
\hline
\hline
$Att$& 0.9690(13)$\cdot$10$^{-6}$ & 0.13\%\\
\hline
\end{tabular}
\end{center}
\caption{Error budget of the attenuation} 
\label{tab:attfactor}
\end{table}

\subsection{Overall stability}
While characterising the detector, the stability of the optical components can affect the uncertainty of the measurement. We measure the Allan deviation of the optical power at D over an extended period ($\sim$18 hours). The power stability of the source is degraded by the testbench but after one hour of measurement the power fluctuations are below 0.02\% of the average value. Because of that, during the efficiency characterisation we check the power every hour with 10\,s of collection time.\\ 
\begin{table}[!h]
\begin{center}
\begin{tabular}{|c|c|c|}
\hline
\multicolumn{3}{|c|}{\bf{Stability of the optical power}}\\
\hline
\bf{Time}&\bf{Stable source}& \bf{point D}\\
\hline
10\,s& 0.8 ppm & 8 ppm\\
\hline
1\,hour&0.6 ppm & 200 ppm\\
\hline
\end{tabular}
\end{center}
\caption{Stability of the optical power} 
\label{tab:stability}
\end{table}

\subsection{Uncertainty budget}
Table\,\ref{tab:setuperrorudget} reports the uncertainty budget measured for the optical power impinging on the detector. The relative uncertainty of the testbench is 0.16\%. Including the uncertainty of the reference power meter, the uncertainty of the number of photons impinging on the detector is 0.59\%. This number is suitable for precise characterisations even at higher detection efficiencies.

\begin{table}[!ht]
\begin{center}
\begin{tabular}{|c|c|c|}
\hline
\multicolumn{3}{|c|}{\bf{Error Budget}}\\
\hline
\bf{Sym.}&\bf{Name}& \bf{Rel. uncertainty (k=1)}\\
\hline
R$_{DC}$& Splitting ratio& 0.09\%\\
\cline{0-2}
$Att$& Attenuation chain & 0.13\%\\
\cline{0-2}
&Stability & 0.02\%\\
\hline
\hline
&{\bf Total} & 0.16\%\\
\hline
\end{tabular}
\end{center}
\caption{Error budget of the testbench.} 
\label{tab:setuperrorudget}
\end{table}

\section{Measurement of the detection efficiency}

The DUT is a pigtailed InGaAs/InP avalanche photodiode with a monolithic integrated negative feedback resistor. The detector is cooled with a Stirling refrigerator down to -70$^{\circ}$C\,\cite{art:Korzh13}. The electronic readout drives the detector in free-running conditions applying a programmable hold\,-off time after each detection to reduce the afterpulsing. 

\begin{figure}[!ht]
\centering\includegraphics[width=8cm]{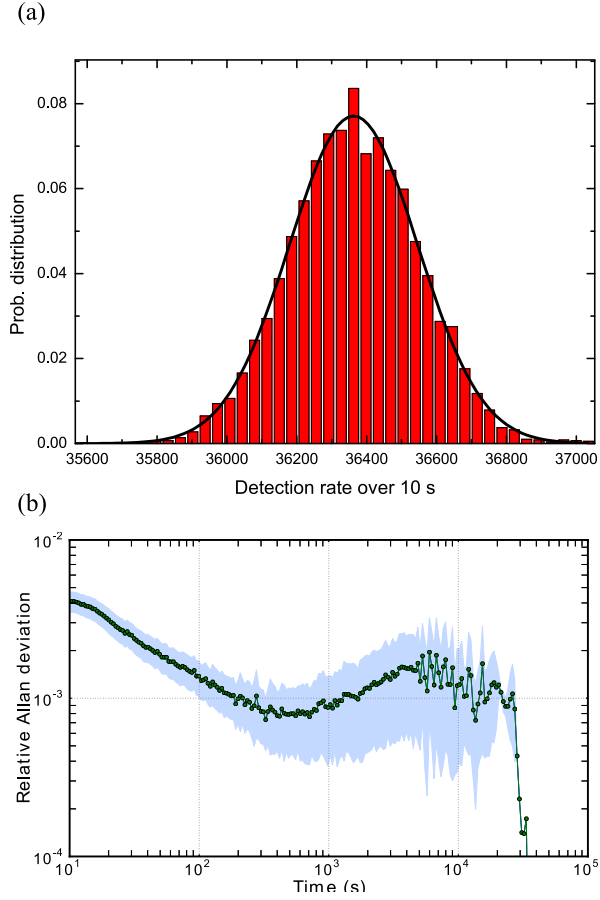}
\caption{(a) Detection rate probability distribution for an ID220 with collection time set to 10\,s. The incoming power is not adjusted during the entire measurement (10 hours). The measured detection rate is limited only by Poissonian statistics and no significant drift of the average value is measured. This demonstrates the high stability of the setup. (b) Relative Allan deviation for the same measurement. The light blue shaded area represents the uncertainty of the measured value.} 
\label{fig:systemstability}
\end{figure}

To quantify the stability of the overall setup, we continuously measure the detection-rate over 10 hours. In Fig.\ref{fig:systemstability}(a) we show the detection-rate probability distribution. Each data point corresponds to 10 seconds of collection time.  The distribution, binned according to Scott's rule\,\cite{art:Scott79}, is Gaussian with standard deviation (184) close to the square root of the average value (191). This is consistent with a Poissonian distribution having a large average value. For such a detection rate (which corresponds to a relative uncertainty smaller than 1\%), the fluctuations of the optical beam are negligible. In Fig.\ref{fig:systemstability}(b) we also show a quantitative analysis by means of the relative Allan deviation for the same dataset. Note that the relative Allan deviation remains below 0.2\% for the entire dataset.

After the calibration of PM$_{\text{ref}}$, to estimate the detection efficiency ($\eta$) of an ideal single-photon detector it would be sufficient to record, with two independent measurements, the avalanche rate sending or not photons to the detector. Then $\eta$ is derived using:
\begin{equation}\label{eq:efficiency}
\eta=\frac{r_{det}-r_{dc}}{N}
\end{equation}
where $r_{det}$ is the avalanche rate originated by both photons and dark counts and $r_{dc}$ is the avalanche rate originated only by dark counts.

However, SPADs are affected by afterpulses which are countered introducing an hold-off time after each detection. This influences the measured rates: 
on the one hand, the detector can not detect photons during the hold-off times so, at a higher count rate, $\eta$ is underestimated. On the other hand, the afterpulses increase the detection rate bringing an overestimation of $\eta$. The impact of the hold-off times on the detection rate can be corrected introducing a duty-cycle corresponding to the time when the detector is on. To correct for the afterpulsing we consider that after any detection an afterpulse is generated with probability p$_{ap}$.
These corrections are introduced modifying $r_{det}$ and $r_{dc}$ with
\begin{eqnarray}
r_{det}(1+p_{ap})=\frac{r_{det}^*}{(1-r_{det}^*\tau)}\label{eq:correction1}\\
r_{dc}(1+p_{ap})=\frac{r_{dc}^*}{(1-r_{dc}^*\tau)}\label{eq:correction2}
\end{eqnarray}
where $\tau$ is the hold-off time and $r_{det}^*$ ($r_{dc}^*$) is the {\bf measured} detection rate (dark-count rate).
We can now plug Eq.\ref{eq:correction1}, \ref{eq:correction2} into Eq.\ref{eq:efficiency} to get\,\cite{art:Lunghi12}:

\begin{equation}\label{eq:etacorr}
\eta = \frac{1}{N\cdot(1+p_{ap})}\left(\frac{r_{det}^*}{1-r_{det}^*\tau}-\frac{r_{dc}^*}{1-r_{dc}^*\tau}\right)
\end{equation}

In our case, $r_{det}^*$, $r_{dc}^*$ have been measured after having integrated the avalanches over an extended period for different hold-off times (5, 10, 20 $\mu$s) and different nominal efficiencies (15\% and 20\%). 
As one may notice, it can happen that an afterpulse is originated by another afterpulse. To consider this, p$_{ap}$ should correspond to the total afterpulsing probability and include also higher-order afterpulses 
(see\,\cite{book:Kindt91}, Sec.6.2.2 Eq. 6.10).\\ 
p$_{ap}$ can be measured reconstructing the temporal evolution of the avalanche rate given that an avalanche has occurred at time zero, $P_{c}(t\mid 0)$. For very long delays there will be no correlation between two events and the probability that a pulse occurs will be determined only by the mean count rate, $n$. We calculate p$_{ap}$ using
\begin{equation}
p_{ap} = \int_0^{75\mu s}(P_{c}(t\mid 0)-n)dt
\end{equation}
We use two independent measurement procedures (see Appendix\,\ref{sec:afterpulse}):
\begin{itemize}
\item The first method, described in\,\cite{book:Kindt91}, records the time interval between successive avalanches. This is the easiest measurement procedure and requires only a Time To Digital Converter device to register the time-stamp. 
\item The second method, described in\,\cite{art:Lunghi12}, uses a pulsed laser to trigger the first avalanche. The temporal evolution is recorded with an FPGA board which controls the setup to record data only when the detector is in a well-defined condition. This method has the advantage of precise control of the detector but it requires a dedicated setup.
\end{itemize}

\subsection{Results}
For our detector, we measure a detection efficiency of 21.2$\pm$0.2\% with a dark-count rate of 89\,Hz at -70$^{\circ}$C. However, here we are interested in the measurement error introduced by an insufficient compensation for hold-off time and the afterpulses. In our case, the uncertainty introduced by the hold-off time correction is negligible (below $\sim$0.04\%) since the length of the hold-off time is known precisely and we pay attention not to saturate the detector (the detector is inactive less than 10\% the time). The afterpulse correction is more delicate as it is illustrated in Tab.\,\ref{tab:correction}. The table reports the efficiencies with different hold-off times before and after correcting for the afterpulsing (for p$_{ap}$ obtained by the 2 methods).

\begin{table}[!ht]
\begin{center}
\begin{tabular}{|c|c|c|c|c||c|c|c||c|c|c|}
\hline
&&\multicolumn{9}{c|}{\bf{Detection efficiency (\%)}}\\
\cline{3-11}
\bf{Bias (V)}&{\bf $r^*_{dc}$(Hz)}&\multicolumn{3}{c||}{\bf{Without $p_{ap}$ corr.}}& \multicolumn{3}{c||}{\bf{1st method}}&\multicolumn{3}{c|}{\bf{2nd method}}\\
\cline{3-11}
&&\bf{5$\mu$s}&\bf{10$\mu$s}&\bf{20$\mu$s}&\bf{5$\mu$s}&\bf{10$\mu$s}&\bf{20$\mu$s}&\bf{5$\mu$s}&\bf{10$\mu$s}&\bf{20$\mu$s}\\
\hline
72.3&49&17.34&16.54&16.20&15.89&16.09&16.06&16.13&16.17&16.09\\
\hline
73.5&89&26.54&22.56&21.55&20.36&20.98&21.19&21.15&21.20&21.27\\
\hline
\end{tabular}
\end{center}
\caption{Detection efficiency estimated before and after applying the afterpulsing correction. For the latter, we compared the two methods used to measure the afterpulsing probability. } 
\label{tab:correction}
\end{table}
One would expect the quantum efficiency of a SPAD to be independent of the chosen hold-off time. This is clearly not the case for the efficiencies obtained without afterpulsing correction. Looking at the values including correction, we note that the differences become much smaller. This is especially the case if P$_{ap}$ is measured with the second method. This indicates that the values obtained with this method are more appropriate. Note that, differences above 0.23\% are significant according to Tab.\ref{tab:efficiencyerrorudget}. However for smaller efficiencies, P$_{ap}$ obtained by the simpler first method can give still satisfactory results, in particular for longer hold-off times. We expect the values for longer hold-off times being a better estimate of the real quantum efficiency. The uncertainty of the quantum efficiency introduced by imperfection of the afterpulses compensation depends on the settings. For 20\% efficiency and 20\,$\mu$s of hold-off time, we assume conservatively the introduced error to be smaller than 0.4\%, which is the difference between the values obtained with 20 $\mu$s and 10 $\mu$s, respectively.
Finally, we can provide the total uncertainty budget for the quantum efficiency characterisation, see Tab.\ref{tab:efficiencyerrorudget} for an overview.
\begin{table}[!ht]
	\begin{center}
		\begin{tabular}{|c|c|c|c|c|}
			\hline
			\multicolumn{5}{|c|}{\bf{Error Budget(Nom. Eff 20\% HO 20$\mu$s)}}\\
			\hline
			\bf{Sym.}&\bf{Name}& \bf{Value}&{\bf{Effect on $\eta$}}&{\bf{Uncertainty(k=1)}}\\
			\hline
			$P_{\text{pm}}$&Power at PM$_{\text{ref}}$ (nW)& 6.59$\pm$0.04& & 0.57\%\\	
			\hline
			&Transmission up to port D&(3.687$\pm$0.006)$\cdot10^{-7}$&0.16\%&\\		
			\cline{0-3}
			$r_{det}$&Detection rate (Hz)&3860$\pm$6& 0.16\%&0.23\%\\
			\cline{0-3}
			$r_{dc}$&Dark-count rate (Hz) &87.84$\pm$0.66&0.02\%&\\
			\hline
			&Corr. for the afterpulses&&0.4\%&0.4\%\\
			\hline
			\hline
			&\bf{Quantum efficiency }&21.19$\pm$0.15\%& &0.73\%\\
			\hline
		\end{tabular}
	\end{center}
	\caption{Error budget of the efficiency characterisation.} 
	\label{tab:efficiencyerrorudget}
\end{table}

\section{Conclusions}
In this paper we have shown how to perform an absolute calibration of the detection efficiency of a fibre-coupled single-photon detector using a simple testbench and only standard optical components. We have determined the detection efficiency of a SPAD with a relative uncertainty well below 1\%. 
If the measurement is performed carefully, correcting for afterpulses and hold-off times, the precision is determined by the absolute calibration of the reference detector. This calibration is achieved using a stable light source and an EDFA radiometer with an error as low as 0.57\%. The EDFA radiometer provides a reliable primary measurement standard and is suitable for calibrating single-photon detectors. 

\section*{Acknowledgments}
The authors thank  A.\,Martin, N.~Walenta, C.~Barreiro, F.~Do Rego Monteiro, T.~Guerreiro for useful discussions. 
We are very grateful to Jacques Morel and Armin Gambon from the Swiss Federal Office for Metrology (METAS) for useful discussions, for hints and ideas, and for calibrating our power meter.
The financial support is provided by the swiss NCCR QSIT.
\newpage
\section*{Appendix}
\appendix

\section{Error analysis of the calibration factor}\label{SM:radio}
Here we provide the analysis for the standard error of a single-shot absolute power measurement using the EDFA radiometer. We assume a flat gain to simplify the analysis.
Equation\,\ref{eq:muin} states
\begin{eqnarray}
\mu_{in}(\lambda) =\left(1-\frac{1}{\overline{G}}\right)\left(\frac{P_{st}^*(\lambda)}{P_{sp}^*(\lambda)}-1\right)\\
\sigma_{\mu_{in}(\lambda)}^2 = \sigma_A^2+\sigma_B^2= A\cdot B
\end{eqnarray}
where
\begin{eqnarray}
\sigma_{A} = \frac{\sigma_{\overline{G}}}{\overline{G}(1-1/\overline{G})}\\
\sigma_{B} = \frac{\Delta (P_{st}^*(\lambda)/P_{sp}^*(\lambda))}{P_{st}^*(\lambda)/P_{sp}^*(\lambda)-1}
\end{eqnarray}
To estimate $\Delta (P_{st}^*(\lambda)/P_{sp}^*(\lambda))$ we repeatedly measure (20 times) $P_{i}^*(\lambda)$ and we calculate its standard error.
The error on $P_{radio}$ can be deduced from Eq.\ref{eq:Pradio} using:
\begin{equation}
\sigma_{P_{radio}}^2 = \sigma_{T_1}^2+\sigma_{T_2}^2+\int (\sigma_{\mu_{in}(\lambda)}^2+\sigma_{\Delta\lambda}^2+\sigma_{\Delta\frac{1}{\lambda^3}}^2)d\lambda
\end{equation}
The calibration factor is 
\begin{eqnarray}
k = \frac{P_{radio}}{P_{in}^*}\\
\sigma_{k}^2 = \sigma_{P_{radio}}^2+\sigma_{P_{in}^*}^2
\end{eqnarray}
where $\sigma_{P_{in}^*}$ is the statistical error on the reading of the P$_{\text{PM}}$.

\begin{table}[!ht]
\begin{center}
\begin{tabular}{|c|c|}
\hline
\multicolumn{2}{|c|}{\bf{Error Budget}}\\
\hline
\bf{Sym.}& \bf{Rel. uncertainty (k=1)}\\
\hline
$\sigma_{\overline{G}}$& 0.26\%\\
\hline
$\frac{P_{st}^*(\lambda)}{P_{sp}^*(\lambda)}$&0.28\%\\
\hline
$\sigma_{\Delta\lambda}$& 0.0002\%\\
\hline
$\sigma_{\Delta\frac{1}{\lambda^3}}$& 0.01\%\\
\hline
$\sigma_{T_1}$&0.24\%\\
\hline
$\sigma_{T_2}$&0.1\%\\
\hline
\bf{$\sigma_{P_{radio}}$}&0.57\%\\
\hline
\bf{$\sigma_{P_{in}^*}$}&0.068\%\\
\hline
\hline
\bf{$\sigma_{k}$}&0.57\%\\
\hline
\end{tabular}
\end{center}
\caption{Error budget of the radiometer.} 
\label{tab:radiometererrorudget}
\end{table}

\section{Afterpulsing characterisation}\label{sec:afterpulse}
As mentioned before, afterpulses are usually characterised starting from the statistical distribution of the time intervals between two avalanches. 
The investigations about the physical nature of the phenomena are usually carried out considering the probability distribution between two {\bf subsequent} avalanches\cite{art:Cova91}. On the contrary, we are interested in having the distribution between {\bf any} pairs of avalanches with no assumption on what is happening between them. The difference between the two distributions is mainly due to higher order afterpulsing. 
We implemented two independent methods able to reconstruct the avalanche rate at time t conditioned on having an avalanche at time zero, $P_c(t\mid\,0)$. 

In the first method\cite{book:Kindt91}, the timestamps of the avalanche occurrences are recorded. Then the histogram of the time delays between two pulses, $h[i\Delta T]$, is built: for every pulse in the timestamp, the delays between this pulse and its successors (until a maximum delay T\,$_{max}=$\,75\,$\mu$s) are calculated and $h[i\Delta T]$ is increased by one for each of these delays. T\,$_{max}$ is chosen considering that the histogram becomes flat when the avalanches are uncorrelated. 
$P_c(t\mid 0)$ is then deduced:
\begin{equation}
P_c(i\Delta T\mid 0)\Delta T = \frac{h[i\Delta T]}{N_{Tot}}
\end{equation} 
where $N_{Tot}$ is the total number of pulses that belong to the timestamp and $\Delta$T is the bin-width, $\sim$\,300\,ns. The main advantage of this method is its simplicity since it requires only a TDC to be performed. However, since the condition of the SPAD before the first avalanche is not well-defined, we can not guarantee that the distribution is independent from the history of the diode.

For this reason, a second method\cite{art:Lunghi12} has been developed: in this method the SPAD is prepared in a well-defined condition, i.e. that no avalanche has occurred in the previous 75$\mu$\,s. Then an FPGA triggers a laser pulse which is sent to the detector. When the pulse is detected, the FPGA records all the avalanches occurring in the next 75$\mu$s, building $h[i\Delta T]$. This time, $N_{Tot}$ corresponds to the total number of avalanches originated by the laser pulses. 

The total afterpulse probabilities, $p_{ap}$, obtained with these methods are reported in Tab.\ref{tab:afterpulse} for different settings of efficiency and hold-off times (HO). For the first method we also estimate the repeatability of the measurement doing the measurement 4 times.
The two methods produce significantly different results. Since the measurement conditions can be well controlled for the second method, we believe that this method gives better results. This hypothesis is reinforced by the analysis of the results in chapter 5.1. However, more work will be necessary to understand the systematic difference between the two methods. It has to be noted that the impact of these discrepancies is small(for  p$_{ap}<<$1) and ,  and finally not limiting the precision of the efficieny measurement. 
\begin{table}[!htbp]
\begin{center}
\begin{tabular}{|c|c|c|c|}
\hline
&\multicolumn{3}{c|}{\bf{15\% Efficiency}}\\
\hline
&\multicolumn{2}{c|}{\bf{Method 1}}&\bf{Method 2}\\
\cline{0-2}
&{\bf Value}&{\bf Repeatability}&\\
\hline
\bf{HO 5$\mu$s} & 9.15(9)\% &1.0\%&7.48\%\\
\hline
\bf{HO 10$\mu$s} & 2.83(4)\% &1.3\%&2.33\%\\
\hline
\bf{HO 20$\mu$s} & 0.83(3)\% &3.6\%& 0.63\% \\
\hline
&\multicolumn{3}{c|}{\bf{20\% Efficiency}}\\
\hline
&\multicolumn{2}{c|}{\bf{Method 1}}&\bf{Method 2}\\
\cline{0-2}
&{\bf Value}&{\bf Repeatability}&\\
\hline
\bf{HO 5$\mu$s} & 30.35(7)\% &0.07\%& 25.51\%\\
\hline
\bf{HO 10$\mu$s} & 7.50(8)\% &1.1\%& 6.40\%\\
\hline
\bf{HO 20$\mu$s} & 1.71(3)\% &1.8\%& 1.34\%\\
\hline
\end{tabular}
\end{center}
\caption{Total afterpulse probabilities. The acquisition time for the first (second) method is 10 minutes (1 hours).} 
\label{tab:afterpulse}
\end{table}
\end{document}